\documentstyle[12pt,epsfig,axodraw,a4]{article} 
\def\bild#1#2{    
        \vspace*{-5mm}
        \begin{center}
        \begin{math}
        \epsfxsize#2cm
        \epsffile{#1}
        \end{math}
        \end{center}  }
\newcommand{\vs}{\vspace{-0.25cm}}
\begin{document} 
\begin{center}
\large{\bf Nuclear energy density functional from chiral pion-nucleon 
dynamics: Isovector spin-orbit terms}

\bigskip

N. Kaiser\\

\medskip

{\small Physik Department T39, Technische Universit\"{a}t M\"{u}nchen,
    D-85747 Garching, Germany}

\end{center}

\medskip

\begin{abstract}
We extend a recent calculation of the nuclear energy density functional in the
systematic framework of chiral perturbation theory by computing the isovector
spin-orbit terms: $(\vec \nabla \rho_p- \vec \nabla \rho_n)\cdot(\vec J_p-\vec
 J_n)\, G_{so}(k_f)+ (\vec J_p-\vec J_n)^2 \, G_J(k_f)$. The calculation 
includes the one-pion exchange Fock diagram and the iterated one-pion exchange
Hartree and Fock diagrams. From these few leading order contributions in the 
small momentum expansion one obtains already a good equation of state of 
isospin-symmetric nuclear matter. We find that the parameterfree results for 
the (density-dependent) strength functions $G_{so}(k_f)$ and $G_J(k_f)$ agree 
fairly well with that of phenomenological Skyrme forces for densities $\rho >
\rho_0/10$. At very low densities a strong variation of the strength functions 
$G_{so}(k_f)$ and $G_J(k_f)$ with density sets in. This has to do with chiral 
singularities $m_\pi^{-1}$ and the presence of two competing small mass scales
$k_f$ and $m_\pi$. The novel density dependencies of $G_{so}(k_f)$ and 
$G_J(k_f)$ as predicted by our parameterfree (leading order) calculation should
be examined in nuclear structure calculations.
\end{abstract}

\bigskip



\bigskip 

\bigskip

Among the various phenomenological interactions that have been used extensively
in the description of nuclei, the Skyrme force \cite{skyrme} has gained much
popularity because of its analytical simplicity and its ability to reproduce 
nuclear properties over the whole periodic table within the self-consistent 
Hartree-Fock approximation. Several Skyrme parameterizations have been tailored
to account for single-particle spectra \cite{sk3}, giant monopole resonances 
\cite{skm} or fission barriers of heavy nuclei \cite{skmstar}. Recently, a new 
Skyrme force which also reproduces the equation of state of pure neutron matter
up to neutron star densities, $\rho_n \simeq 1.5 \,{\rm fm}^{-3}$, has been 
proposed in ref.\cite{sly} for the study of nuclei far from stability. A 
microscopic interpretation of the various parameters entering the effective 
Skyrme forces is generally put aside. Sometimes the energy density functional 
is just parameterized without reference to any effective (zero-range) 
NN-interaction \cite{reinhard}.   

Another widely and successfully used approach to nuclear structure calculations
are relativistic mean-field models \cite{walecka,ringreview}. In these 
models the nucleus is described as a collection of independent Dirac-particles 
moving in self-consistently generated scalar and vector mean-fields. The 
footprints of relativity become visible through the large nuclear spin-orbit 
interaction which emerges in that framework naturally from the interplay of the
two strong and counteracting (scalar and vector) mean-fields. The corresponding
many-body calculations are usually carried out in the Hartree approximation,
ignoring the negative-energy Dirac-sea. For a recent review on self-consistent 
mean-field models for nuclear structure, see ref.\cite{reinhard}. In that 
article the relationship between the relativistic mean-field models and the 
Skyrme phenomenology is also discussed. 

The first necessary conditions to be fulfilled by any phenomenological 
nucleon-nucleon interaction come from the (few empirically known) properties of
infinite nuclear matter. These are the saturation density $\rho_0=2k_{f0}^3/3
\pi^2$, the binding energy per particle $-\bar E(k_{f0})$ and the compression
modulus $K =k_{f0}^2 \bar E''(k_{f0})$ of isospin-symmetric nuclear matter as
well as the asymmetry energy $A(k_{f0})$. In a recent work \cite{nucmat}, we
have applied the systematic framework of chiral perturbation theory to the 
nuclear matter many-body problem. In this calculation the
contributions to the energy per particle, $\bar E(k_f)$, originate exclusively
from one- and two-pion exchange between nucleons and they are ordered in powers
of the Fermi-momentum $k_f$ (modulo functions of $k_f/m_\pi$). It has been 
demonstrated in ref.\cite{nucmat} that the empirical saturation point $(\rho_0 
\simeq 0.17\,{\rm fm}^{-3}\,, \bar E(k_{f0})\simeq -15.3\,{\rm MeV})$ and the 
nuclear matter compressibility $K\simeq 255\,$MeV can be well reproduced at 
order ${\cal O}(k_f^5)$ in the small momentum expansion with just one single 
momentum cut-off scale of $\Lambda \simeq 0.65 \,$GeV which parameterizes the 
short-range NN-dynamics necessary for nuclear binding. Most surprisingly, the 
prediction for the asymmetry energy, $A(k_{f0})=33.8\,$MeV \cite{nucmat}, is in
good agreement with its empirical value. In fact very similar results for
nuclear matter can be obtained already at order ${\cal O}(k_f^4)$ in the small
momentum expansion (by dropping the relativistic $1/M^2$-correction to
$1\pi$-exchange and the irreducible $2\pi$-exchange of order ${\cal O}(k_f^5)$)
with a somewhat reduced cut-off scale of $\Lambda \simeq 0.61\,$GeV (for 
detailed results see ref.\cite{lutzpot}). 

Given the fact that many properties of nuclear matter can be well described by
chiral $\pi N$-dynamics treated perturbatively up to three-loop order it is
natural to consider in a further step the energy density functional relevant
for inhomogeneous many-nucleon systems (i.e. finite nuclei). Such an extension
to inhomogeneous many-nucleon systems can be done with the help of the density 
matrix-expansion of Negele and Vautherin \cite{negele}. The bilocal 
density-matrix (given by a sum over the occupied energy eigenfunctions) is
expanded in relative and center-of-mass coordinates with expansion coefficients
determined by purely local quantities: nucleon density $\rho(\vec r\,)$, 
kinetic energy density $\tau(\vec r\,)$ and spin-orbit density $\vec J(\vec r\,
)$. The Fourier-transform of the (so expanded) density-matrix defines in 
momentum-space a "medium-insertion" $\Gamma(\vec p,\vec q\,)$ for the 
inhomogeneous many-nucleon system which then allows to calculate 
diagrammatically the nuclear energy density functional.  

In a recent work \cite{efun} we have considered the isospin-symmetric case of 
equal (even) proton and neutron number. The corresponding energy 
density functional takes the general form: ${\cal E}[\rho,\tau,\vec J\,] =\rho
\bar E(k_f)+[\tau-3\rho k_f^2/5]/2\widetilde M^*(\rho)+ (\vec \nabla \rho)^2\, 
F_\nabla(k_f)+  \vec \nabla \rho \cdot\vec J\, F_{so}(k_f)+ \vec J\,^2 \, 
F_J(k_f)$. We have found that the effective nucleon mass $\widetilde M^*(\rho)$
deviates at most by $\pm15\%$ from its free space value $M$, with $0.89M<
\widetilde M^*(\rho)<M$ for $\rho < 0.11 \,{\rm fm}^{-3}$ and $\widetilde M^*
(\rho)>M$ for higher densities $\rho<\rho_0 = 0.174\,{\rm fm}^{-3}$. 
Interestingly, a recent large scale fit of (almost two thousand) nuclide masses
by Pearson et al. \cite{pearson} finds a similarly  enhanced effective nucleon
mass: $\widetilde M^*(\rho_0) \simeq 1.05M$. The strength of the $(\vec \nabla 
\rho)^2$-term, $F_\nabla(k_{f0})$, is (at saturation density) comparable to
that of phenomenological Skyrme forces. The magnitude of $F_J(k_{f0})$
accompanying the squared spin-orbit density $\vec J\,^2$ comes out considerably
larger. Both quantities increase strongly as the nucleon density $\rho$ tends
to zero. This has to do with the explicit presence of the small mass scale
$m_\pi=135\,$MeV which amounts to about half of the Fermi momentum $k_{f0}$ in 
equilibrated nuclear matter. The strength of the isoscalar nuclear spin-orbit 
interaction, $F_{so}(k_{f0})$, as given by iterated $1\pi$-exchange is (at 
saturation density) about half as large as the corresponding empirical value
$\sim 90\, $MeVfm$^5$, however, with the wrong negative sign. This isoscalar 
spin-orbit interaction is not a relativistic effect but proportional to the 
large nucleon $M$. From that result it becomes clear that perturbative chiral 
pion-nucleon dynamics (alone) cannot account for the mechanisms underlying the
empirical isoscalar nuclear spin-orbit force, whereas relativistic
scalar-vector mean-field models \cite{walecka,ringreview} give a successful
phenomenology of it. 

The purpose of the present work is to calculate using the same framework as in
ref.\cite{efun} the isovector spin-orbit interactions generated by one- and
two-pion exchange in order to reveal the underlying isospin dependence. Note,
for example, that the Skyrme force gives rise to  isoscalar and isovector 
spin-orbit interactions with a fixed ratio 3:1. In particular, this 
restrictive feature of the Skyrme energy density functional has been made 
responsible for the less accurate description of isotope shifts in the Pb 
region \cite{isoorbit} in comparison to relativistic mean-field calculations. 
Naturally, one expects that the finite-range character of the two-pion exchange
interaction will lift such restrictive features of the zero-range Skyrme force.

Let us begin with writing down the explicit form of the isovector spin-orbit 
terms in the nuclear energy density functional:
\begin{equation} {\cal E}[\rho_p,\rho_n,\vec J_p,\vec J_n] =  (\vec \nabla 
\rho_p- \vec \nabla \rho_n)\cdot(\vec J_p-\vec J_n)\, G_{so}(k_f)+ (\vec J_p-
\vec J_n)^2 \, G_J(k_f)+ \dots \end{equation} 
Here, 
\begin{equation} \rho_{p,n}(\vec r\,) = {k_{p,n}^3(\vec r\,) \over 3\pi^2} = 
\sum_{\alpha \in occ} \Psi^{(\alpha)
\dagger}_{p,n}( \vec r\,)\Psi^{(\alpha)}_{p,n}( \vec r\,)\,,\end{equation} 
are the local proton and neutron densities written in terms of the 
corresponding (local) proton and neutron Fermi-momenta $k_{p,n}(\vec r\,)$, and
expressed as sums over the occupied single-particle orbitals $\Psi^{(\alpha)
}_{p,n}( \vec r\,)$. The spin-orbit densities of protons and neutrons are 
defined similarly: 
\begin{equation} \vec J_{p,n}(\vec r\,) = \sum_{\alpha \in occ} \Psi^{(\alpha)
\dagger}_{p,n}(\vec r\,)i\, \vec \sigma \times \vec \nabla\Psi^{(\alpha)}_{p,n
}( \vec r\,) \,. \end{equation} 
Furthermore, $G_{so}(k_f)$ and $G_J(k_f)$ denote the associated strength
functions. In Skyrme parameterizations \cite{sk3,skm,skmstar,sly,pearson} these
are just constants, $G_{so}(k_f) = W_0/4$ and $G_J(k_f) = (t_1-t_2)/32$,
whereas in our calculation their explicit density dependence originates from
the finite-range character of the $1\pi$- and $2\pi$-exchange interaction.

The starting point for the construction of an explicit nuclear energy density 
functional ${\cal E}[\rho_p,\rho_n,\vec J_p,\vec J_n]$ is the bilocal 
density-matrix as given by a sum over the occupied energy eigenfunctions: 
$\sum_{\alpha\in occ}\Psi^{(\alpha)}_{p,n}( \vec r -\vec a/2)\Psi^{(\alpha)
\dagger}_{p,n}(\vec r +\vec a/2)$. According to Negele and Vautherin
\cite{negele} it can be expanded in relative and center-of-mass coordinates,
$\vec a$  and $\vec r$, with expansion coefficients determined by local
quantities (nucleon density, kinetic energy density, spin-orbit density). As
outlined in section 2 of ref.\cite{efun} the Fourier-transform of the (so
expanded) density-matrix defines in momentum-space a medium-insertion
$\Gamma(\vec p,\vec q\,)$ for the inhomogeneous many-nucleon system. It is
straightforward to generalize this construction to the isospin-asymmetric 
situation of different proton and neutron local densities $\rho_{p,n}(\vec 
r\,)$ and $\vec J_{p,n}(\vec r\,)$. We display here only that part of the
medium-insertion $\Gamma(\vec p,\vec q\,)$ which is actually relevant for the
diagrammatic calculation of the isovector spin-orbit terms eq.(1): 
\begin{eqnarray} \Gamma(\vec p,\vec q\,)& =& \int d^3 r \, e^{-i \vec q \cdot
\vec r}\,\bigg\{ {1+\tau_3 \over 2}\theta(k_p-|\vec p\,|) +{1-\tau_3 \over 2}
\theta(k_n-|\vec p\,|) \nonumber \\ && +{\pi^2 \over 4k_f^4}
\Big[\delta(k_f-|\vec p\,|) -k_f \,\delta'(k_f-|\vec p\,|) \Big]\, \tau_3 \, 
\vec \sigma\cdot [\vec p \times (\vec J_p-\vec J_n)] +\dots  \bigg\}\,.  
\end{eqnarray}
When working to quadratic order in deviations from isospin symmetry
(proton-neutron differences) it is sufficient to use an average Fermi-momentum 
$k_f$ in the prefactor of the isovector spin-orbit density $\vec J_p-\vec J_n$.
The double line in the left picture of Fig.\,1 symbolizes this medium insertion
together with the assignment of the out- and in-going nucleon momenta $\vec p 
\pm \vec q/2$. The momentum transfer $\vec q$ is provided by the Fourier 
components of the inhomogeneous (matter) distributions $\rho_{p,n}(\vec r\,)$ 
and $\vec J_{p,n}(\vec r\,)$.

\bigskip

\bild{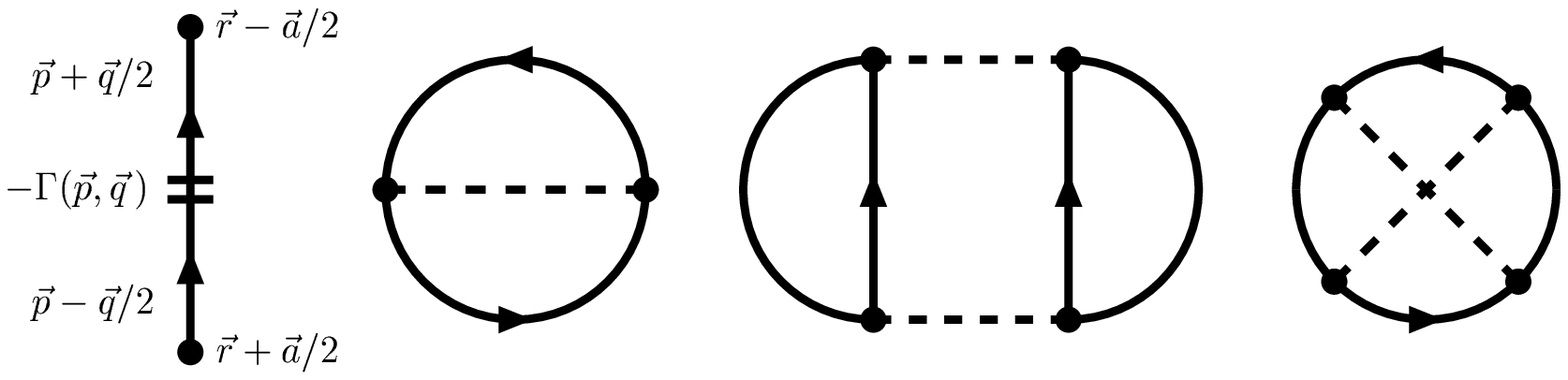}{12}
{\it Fig.\,1: Left: The double line symbolizes the medium insertion defined by
eq.(4). Next are shown: The two-loop one-pion exchange Fock-diagram and the
three-loop iterated one-pion exchange Hartree- and Fock-diagrams. Their
combinatoric factors are 1/2, 1/4 and 1/4, in the order shown.}
\bigskip

Now we turn to the analytical evaluation of the pion-exchange diagrams shown in
Fig.\,1. We give for each diagram only the final result omitting all technical
details related to extensive algebraic manipulations and solving elementary 
integrals. A collection of the relevant "master integrals" can be found in the
appendix of ref.\cite{efun}. We obtain from the $1\pi$-exchange Fock diagram 
with two medium insertions: 
\begin{equation} G_J(k_f) = {g_A^2\over(8 m_\pi f_\pi)^2} \bigg\{-{10+24u^2 
\over 3(1+4u^2)^2}-{1\over 6u^2} \ln(1+4u^2) \bigg\} \,, \qquad u={k_f \over 
m_\pi} \,. \end{equation}
This contribution to $G_J(k_f)$ is just $-1/3$ of the contribution to the 
isoscalar strength function $F_J(k_f)$ (see eq.(10) in ref.\cite{efun}) as a 
simple consequence of the isospin-trace tr$[\tau_a (J_s+\tau_3 J_v)\tau_a (J_s
+\tau_3 J_v)]= 6 J_s^2 - 2J_v^2$. There is no contribution of the
$1\pi$-exchange Fock-diagram to the strength function $G_{so}(k_f)$ since the 
spin-trace with one insertion proportional to  $\vec \sigma \cdot 
(\vec p \times \vec J_v)$ vanishes. We note also that the $1\pi$-exchange
Hartree diagram (not shown in Fig.\,1) vanishes identically either because of a
zero spin-trace or because of the momentum integral $\int d^3 p \,\vec p \,
\theta(k_{p,n}-|\vec p\,|) = \vec 0$. The iterated one-pion exchange Hartree 
diagram with two medium insertions leads to the result:
\begin{equation} G_{so}(k_f) = {g_A^4M \over \pi m_\pi(4f_\pi)^4} \bigg\{
{1 \over u^2} \ln(1+4u^2)-{8\over 3(1+4u^2)} \bigg\} \,,  \end{equation}
which is $-2/3$ of the contribution to the isoscalar spin-orbit strength
$F_{so}(k_f)$ (see eq.(13) in ref.\cite{efun}). Let us briefly explain the
mechanism which generates the strength function $G_{so}(k_f)$. The exchanged 
pion-pair transfers the momentum $\vec q$ between the left and the right 
nucleon ring and this momentum $\vec q$ enters also the pseudovector 
$\pi N$-interaction vertices. The isovector spin-orbit strength $G_{so}(k_f)$ 
arises from the spin-trace tr$[\vec \sigma \cdot (\vec l+\vec q/2)\,\vec\sigma
\cdot (\vec l -\vec q/2)\,\vec \sigma \cdot (\vec p\times \vec J_v\,)] =
2i\,(\vec q  \times \vec l\,)\cdot(\vec p \times \vec J_v\,)$ where 
$\vec q$ gets converted to $\vec \nabla k_p-\vec \nabla k_n \simeq \pi^2(\vec
\nabla \rho_p-\vec \nabla \rho_n)/k_f^2$ by Fourier transformation. From the 
iterated one-pion exchange Fock diagram with two medium insertions we obtain:
\begin{eqnarray} G_{so}(k_f) &=& {5g_A^4M \over 6\pi m_\pi(4f_\pi)^4} \bigg\{
 {\arctan 2u \over u(1+2u^2)}-{3+4u^2 \over u(1+2u^2)}\arctan u 
\nonumber \\ && +{1\over 2u^2}\ln{1+u^2\over 1+4u^2}+\int_0^u \!dx
{\arctan 2x-\arctan x\over u^2(1+2x^2)} \bigg\} \,,  \end{eqnarray}
\begin{eqnarray} G_J(k_f) &=& {5g_A^4M \over 3\pi m_\pi(8f_\pi)^4} \bigg\{
{2(5+8u^2)\over u(1+2u^2)}\arctan u -{2 \arctan 2u \over u(1+2u^2)}+{1\over 
1+u^2}\nonumber \\ &&+{1\over u^2}\ln{1+4u^2\over 1+u^2}+{2\over u^2}\int_0^u
\!dx {\arctan x-\arctan 2x\over 1+2x^2} \bigg\} \,.  \end{eqnarray}
One notices a relative factor of $-5/3$ in comparison to the contributions to
the isoscalar strength functions $F_{so,J}(k_f)$ (see eqs.(16,17) in
ref.\cite{efun}) which comes from the isospin-trace tr$[\tau_a (J_s+\tau_3 J_v)
\tau_b\tau_a (J_s+\tau_3J_v)\tau_b]= 10 J_v^2 - 6J_s^2$ of that diagram. In
case of the iterated one-pion exchange Hartree diagram with three medium 
insertions one has to evaluate nine-dimensional principal value integrals
over the product of three Fermi spheres. We find the following contribution to 
the isovector spin-orbit strength:
\begin{eqnarray} G_{so}(k_f) &=& {4g_A^4 M u^{-6}\over 3\pi^2 m_\pi(4 f_\pi)^4}
\int_0^u \!dx x^2\int_{-1}^1 \!dy \bigg\{{2u s^4 [2 x^2y^2(s-s')+u^2(2s'-s)] 
\over (1+s^2)^2 (u^2-x^2y^2)}\nonumber \\ && + \bigg[{u(3u^2-5x^2y^2)\over u^2-
x^2y^2} -4xy H \bigg] \bigg[ 3 \arctan s-{3s+2s^3\over 1+s^2}\bigg] \nonumber 
\\ &&  + {H s^3 \over (1+s^2)^3}\Big[(8xy-5s-s^3)s'^2-2xys'(7s+3s^3) \nonumber 
\\ && +s^2xy (11+7s^2)+(s+s^3)(2xy-s)(s''-s') \Big]\bigg\}\,.   \end{eqnarray}
with the auxiliary functions $H=\ln(u+xy)-\ln(u-xy)$ and $s=xy +\sqrt{u^2-x^2+
x^2y^2}$. The partial derivatives of $s$ are abbreviated by $s'=u \partial s/
\partial u$ and $s''=u^2 \partial^2 s/\partial u^2$. In comparison to the
contribution to the isoscalar spin-orbit strength $F_{so}(k_f)$ (see eq.(20) in
ref.\cite{efun}) relative isospin factors of $\pm 2/3$ have occurred in various
subparts. The contribution of the iterated one-pion exchange Hartree diagram
with three medium insertions to the strength function $G_J(k_f)$ reads: 
\begin{eqnarray} G_J(k_f) &=& {g_A^4 M \over 3\pi^2 m_\pi(4  f_\pi)^4} \bigg\{
{1+12u^2-24u^4-96u^6 \over u(1+4u^2)^3} -{1+2u^2 \over 4u^3} \ln(1+4u^2)
\nonumber \\ &&  +\int_0^1 \!dy\, {8u^3 y^2 \over (1+4u^2y^2)^4} \Big[5- (30+
32u^2)y^2\nonumber \\ && +(35+24u^2-16u^4)y^4+56u^2y^6+48 u^4 y^8\Big]\ln{1+y 
\over 1-y} \bigg\} \,. \end{eqnarray}
It is obtained if only if both insertions proportional to $\vec \sigma \cdot 
(\vec  p_{1,2}\times \vec J_v\,)$ stand under a single spin-trace and this 
feature explains also the relative isospin factor $-1/3$ in comparison to 
the contribution to $F_J(k_f)$ written in eq.(21) of ref.\cite{efun}. The 
iterated one-pion exchange Fock diagram with three medium insertions is most 
tedious to evaluate. It is advisable to split the contributions to the strength
functions $G_{so,J}(k_f)$ into "factorizable" and "non-factorizable" parts. 
These two pieces are distinguished by whether the nucleon propagator in the 
denominator can be canceled or not by terms from the product of $\pi 
N$-interaction vertices in the numerator. We find the following "factorizable"
contributions:  
\begin{eqnarray} G_{so}(k_f) &=& {g_A^4 M u^{-3} \over 3\pi^2 m_\pi(4f_\pi)^4} 
\Bigg\{{(1-2u^2)(2u^4-4u^2-1) \over (1+u^2)(1+4u^2)}\ln(1+4u^2)- {1+2u^2\over 
32u^2} \ln^2(1+4u^2)\nonumber \\ &&  -{u^2(11+52u^2) \over 2(1+4u^2)} 
+{u(5+39u^2+64u^4) \over (1+u^2)(1+4u^2)} \arctan 2u + 5\int_0^u \!dx\,\bigg\{
-3u^3x^{-2} \nonumber \\ &&+\Big[4x^2+1+u^2-3x^{-2}(1+u^2)^2\Big]u L^2+\Big[6(
u^2+u^4)x^{-2} +2\nonumber \\ &&-(1+5u^2+5ux)[1+(u+x)^2]^{-1}-(1+5u^2-5ux)[1+ 
(u-x)^2]^{-1}\Big]L\bigg\}  \Bigg\} \,,\end{eqnarray}
\begin{eqnarray} G_J(k_f) &=& {g_A^4 M u^{-3} \over 3\pi^2 m_\pi (8f_\pi)^4} 
\Bigg\{{11-32u^2-8u^4\over u^2(1+u^2)}\ln(1+4u^2) -{3+12u^2+8u^4 \over 4u^4} 
\ln^2(1+4u^2) \nonumber \\ && +{10(2u^2-5)(2+3u^2) \over u(1+u^2)} \arctan 2u 
+{8(248u^6+1102u^4+509u^2+63) \over 3(1+4u^2)^2} \\ && +40 \int_0^u \!dx\,
\bigg\{\Big[2+2u^2-3(1+u^2)^2 x^{-2}-3x^2\Big]u L^2 +2\Big[3x^{-2}(u^2+u^4)-2
-u^2 \nonumber \\ && +(1+ux-u^2)[1+(u+x)^2]^{-1} + (1-ux-u^2)
[1+(u-x)^2]^{-1}\Big]L- 3u^3x^{-2}\bigg\} \Bigg\} \nonumber \,, \end{eqnarray}
with the auxiliary function:
\begin{equation} L= {1\over 4x} \ln{1+(u+x)^2\over 1+(u-x)^2} \,.\end{equation}
The non-factorizable contributions from the iterated one-pion exchange Fock 
diagram with three medium insertions (stemming from nine-dimensional principal 
value integrals over the product of three Fermi spheres) read on the other
hand:  
\begin{eqnarray} G_{so}(k_f) &=& {g_A^4 M \over 3\pi^2 m_\pi(4f_\pi)^4}
\int_{-1}^1\!dy \int_{-1}^1 \!dz {yz \,\theta(y^2+z^2-1)  \over |yz|\sqrt{y^2+
z^2-1}}\bigg\{ 16y^2z\,\theta(y)\theta(z) \bigg[{1+2u^2y^2 \over (1+4u^2y^2)^2}
\nonumber \\ && \times \Big(2uz -\arctan 2uz\Big) +{ u^3z(2z^2-1)\over
(1+4u^2y^2)(1+4u^2 z^2)} \bigg]+\int_0^u \!dx \, {5u^{-8}x^2 st^2 t'\over 2 
(1+s^2)^2(1+t^2)} \nonumber \\ && \times \Big[(s+s^3) (s''-s')(st+sxz -txy) 
+s'^2((3s+s^3)(t+xz)-2txy ) \Big]\bigg\}\,, \end{eqnarray} 
\begin{eqnarray} G_J(k_f) &=& {g_A^4 M \over 3\pi^2 m_\pi(4f_\pi)^4}
\int_{-1}^1\!dy \int_{-1}^1 \!dz {yz \,\theta(y^2+z^2-1) 
\over |yz|\sqrt{y^2+z^2-1}}\bigg\{ y^2\, \theta(y)\theta(z) \bigg[
\Big[4u^2 z^2-\ln(1+4u^2 z^2) \Big] \nonumber \\ && \times {9 +4u^2(5+2y^2)
+16u^4 (y^2+y^4) \over u(1+4u^2y^2)^3} +{ 16u^3(3+4u^2y^2)z^2 (2z^2-1) \over  
(1+4u^2y^2)^2(1+4u^2 z^2)} \bigg]\nonumber \\ &&+\int_0^u \!dx \,
{5x^4 s^2t^2(y^2+z^2-1) \over 4u^{10} (1+s^2)^2(1+t^2)^2}\Big[(s+s^3)(s''-s')+
(3+s^2)s'^2\Big]\nonumber \\ && \times \Big[(t+t^3)(t''-t')+ (3+t^2)t'^2\Big]
\bigg\}\,, \end{eqnarray}
with the auxiliary function $t=xz +\sqrt{u^2-x^2+x^2z^2}$ and its partial
derivatives $t'=u \partial t/\partial u$ and $t''=u^2 \partial^2 t/\partial 
u^2$. In comparison to the contributions to the isoscalar strength functions
$F_{so,J}(k_f)$ (see eqs.(25,26,29,30) in ref.\cite{efun}) there occur relative
isospin factors of $-1/3$ and $-5/3$ in various subparts. Let us add some 
general power counting considerations for the nuclear energy density functional
${\cal E}[\rho_p,\rho_n,\vec J_p,\vec J_n]$. Counting the Fermi momenta 
$k_{p,n,f}$, the pion mass $m_\pi$ and a spatial gradient $\vec \nabla$ 
collectively as small momenta one deduces from eqs.(2,3) that the nucleon 
densities $\rho_{p,n}(\vec r\,)$ and the spin-orbit densities $\vec J_{p,n}(
\vec r\,)$ are quantities of third and fourth order in small momenta,
respectively. With these counting rules the contributions from $1\pi$-exchange
to the nuclear energy density functional ${\cal E}[\rho_p,\rho_n,\vec J_p,\vec 
J_n]$ are of sixth order in small momenta while all contributions from iterated
$1\pi $-exchange are of seventh order. Concerning NN-interactions induced by 
pion-exchange the isovector spin-orbit terms presented here are in fact 
complete up-to-and-including seventh order in small momenta.

Next we turn to the numerical results. We use the physical parameters: $M= 939
\,$MeV (nucleon mass), $m_\pi = 135\,$MeV (neutral pion mass), $f_\pi = 92.4\,
$MeV (pion decay constant) and $g_A = 1.3$ (equivalent to a $\pi N$-coupling
constant $g_{\pi N} = g_A M/f_\pi = 13.2)$. The full line in Fig.\,2 shows the 
result of iterated $1\pi$-exchange for the strength function $G_{so}(k_f)$ 
belonging to the isovector spin-orbit coupling term $(\vec\nabla \rho_p-\vec
\nabla \rho_n)\cdot(\vec J_p-\vec J_n)$ versus the nucleon density $\rho =
2k_f^3/3\pi^2$. For comparison we have drawn the constant values $G_{so}(k_f)=
W_0/4$ of the three Skyrme forces Sly \cite{sly}, MSk \cite{pearson} and SkP 
\cite{skp} (horizontal dashed lines). In the case of Sly and MSk we have 
performed averages over the various parameter sets Sly4-7 and MSk1-6. One
observes a fair agreement of our parameterfree prediction with these empirical
values. In contrast to the isoscalar spin-orbit strength $F_{so}(k_f)$ (see
Fig.\,4 in ref.\cite{efun}) iterated $1\pi$-exchange gives now the correct
positive sign. The density dependence of the isovector spin-orbit strength 
$G_{so}(k_f)$ is moderate for densities $\rho > \rho_0/10)$. A rapid decrease 
sets however in when $\rho$ tends to zero. 

\bigskip
\bild{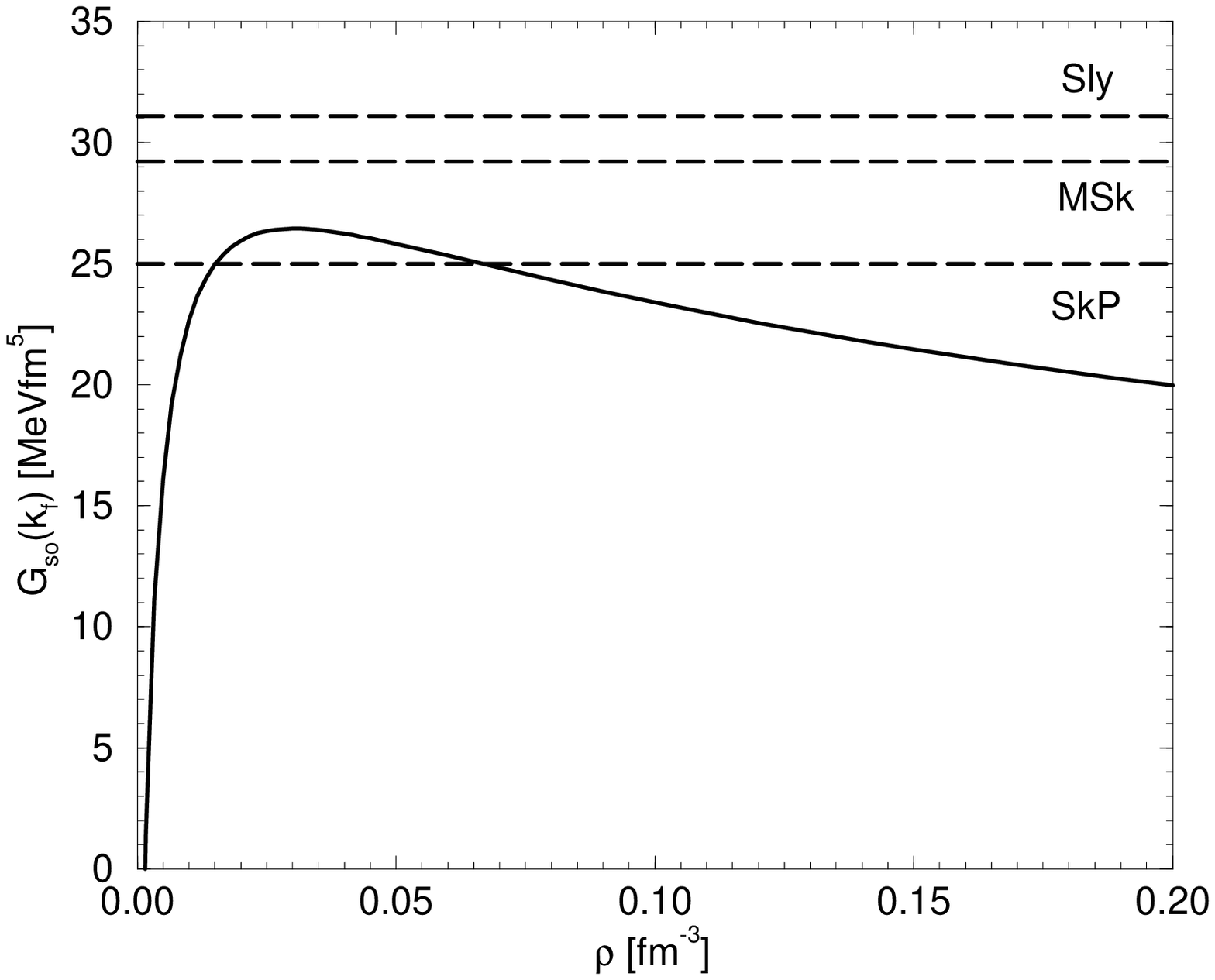}{14}
\vspace{-1.3cm}
{\it Fig.\,2: The strength function $G_{so}(k_f)$ related to the isovector
spin-orbit term $(\vec\nabla \rho_p-\vec\nabla \rho_n)\cdot(\vec J_p-\vec J_n)$
in the nuclear energy density functional versus the nucleon density $\rho=
2k_f^3/3\pi^2$. The three horizontal dashed lines show the constant values 
$G_{so}(k_f)=W_0/4$ of the Skyrme forces Sly \cite{sly}, MSk \cite{pearson} and
SkP \cite{skp}.} 
\bigskip

Finally, we show in Fig.\,3 the strength function $G_J(k_f)$ accompanying the 
squared isovector spin-orbit density $(\vec J_p-\vec J_n)^2$ in the nuclear 
energy density functional versus the nucleon density $\rho=2k_f^3/3\pi^2$.  For
comparison we have drawn the constant values $G_J(k_f)=(t_1-t_2)/32$ of the 
three Skyrme forces Sly \cite{sly}, SkP \cite{sly} and MSk \cite{pearson}, 
(dashed lines). One observes that our parameterfree prediction for the strength
function $G_J(k_f)$ lies well within the band spanned by these empirical
values. Again, the density dependence of $G_J(k_f)$ is moderate for densities 
$\rho>\rho_0/5$ and a rapid decrease sets in when $\rho$ tends to zero. 
Although not visible each (full) curve in Figs.\,2,3, approaches a finite 
(negative) value at  $\rho=0$. One can analytically derive the following low
density limits: 
\begin{equation} G_{so}(0) = -{g_A^4 M \over 3\pi m_\pi (4f_\pi)^4} = -33.8
{\rm MeVfm}^5 \,,  \end{equation}
\begin{equation} G_J(0) = {g_A^2 \over (4m_\pi f_\pi)^2} \bigg[-1+{15 
g_A^2  M m_\pi \over 256 \pi f_\pi^2} \bigg] = -108.0\,{\rm MeVfm}^5 \,, 
\end{equation}
to which only the diagrams with two medium insertions contribute. The large 
numbers in eqs.(16,17) arise from negative powers of the pion mass $m_\pi$ 
(so-called chiral singularities). The most singular $m_\pi^{-2}$-term can be
traced back to the $1\pi$-exchange Fock diagram. Note also the relation between
the isovector and isoscalar spin-orbit strengths at zero density, $G_{so}(0) = 
F_{so}(0)/3$ (for $F_{so}(0)$ see eq.(42) in ref.\cite{efun}). This is a 
necessary consistency check on our diagrammatic calculation. At extremely low 
densities ($k_f <<m_\pi/2$) even the pion-exchange interaction becomes 
effectively short-ranged and therefore the constraint $G_{so}(k_f) =
F_{so}(k_f)/3$ known from the (zero-range) Skyrme spin-orbit force must hold. 

In this context is important to keep in mind that if pionic degrees of
freedom are treated explicitly in the nuclear matter problem the low density
limit is realized only at extremely low densities $k_f<<m_\pi/2$. Often, the
opposite limit where the pion mass $m_\pi$ can be neglected against the Fermi
momentum $k_f$ is already applicable at the moderate densities relevant for
conventional nuclear physics. This is exemplified here by the approximate
density dependence $G_{so,J}(k_f)\sim k_f^{-1}$ (see Figs.\,2,3). Such a 
$\rho^{-1/3}$-behavior becomes exact in the chiral limit $m_\pi=0$ as can be 
deduced by simple mass dimension counting of the iterated $1\pi$-exchange
diagrams (the basic argument is that $M/f_\pi^4 k_f$ has the correct unit of
MeVfm$^5$). 

\bigskip

\bild{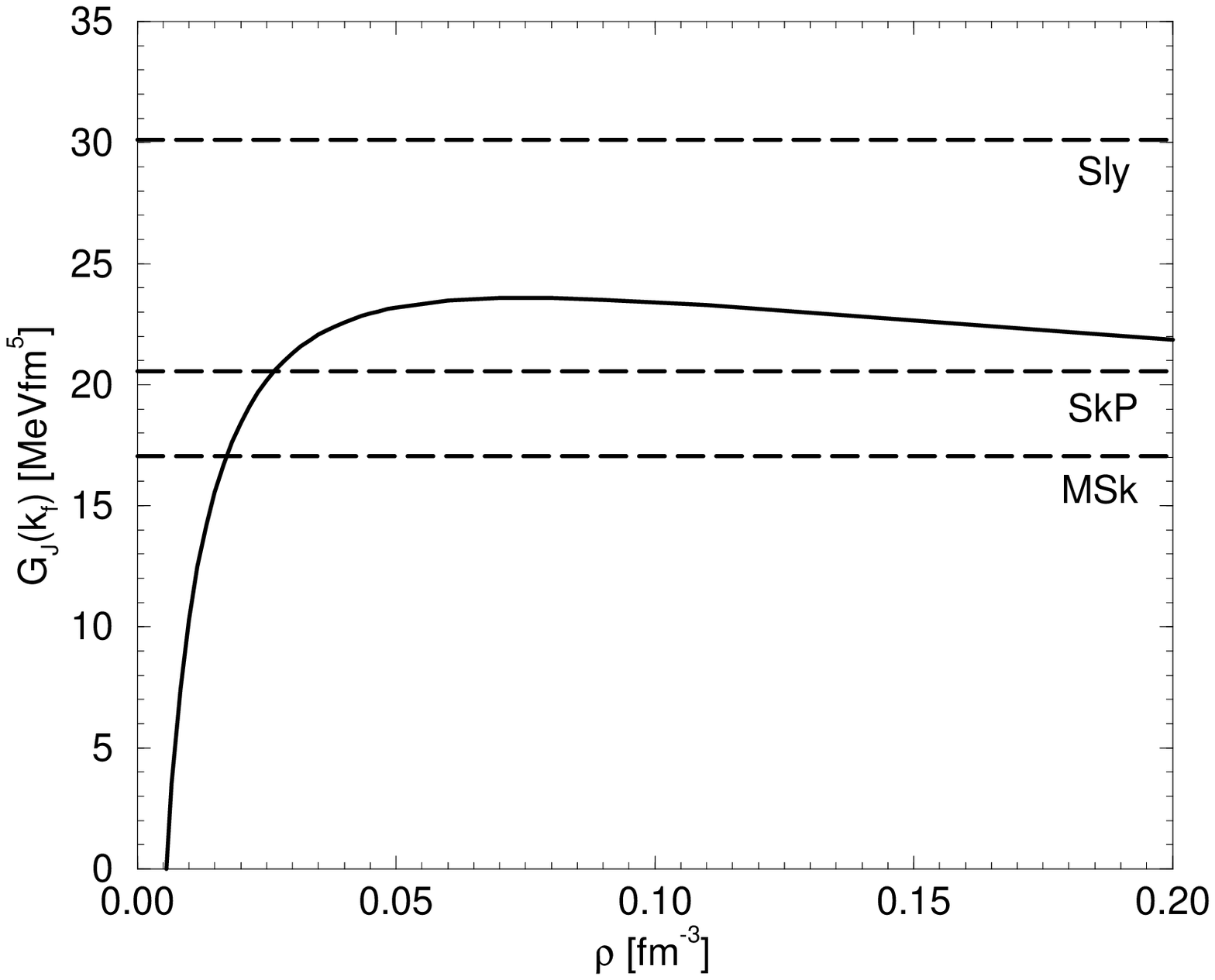}{14}
\vspace{-1.3cm}
{\it Fig.\,3: The strength function $G_J(k_f)$ accompanying the squared 
isovector spin-orbit density $(\vec J_p-\vec J_n)^2$ in the nuclear energy 
density functional versus the nucleon density $\rho=2k_f^3/3\pi^2$. The three 
horizontal dashed lines show the constant values $G_J(k_f)=(t_1-t_2)/32$ of the
Skyrme forces Sly \cite{sly}, SkP \cite{skp} and MSk \cite{pearson}.} 
\bigskip

In summary we extended in this work our recent calculation of the nuclear
energy density functional \cite{efun} in the framework of chiral perturbation
by computing the isovector spin-orbit terms: $(\vec \nabla \rho_p- \vec \nabla
\rho_n)\cdot(\vec J_p-\vec J_n)\, G_{so}(k_f)+(\vec J_p-\vec J_n)^2\,G_J(k_f)$.
Our calculation includes the $1\pi$-exchange Fock diagram and the iterated
$1\pi$-exchange Hartree and Fock diagrams. These few leading order
contributions in the small momentum expansion give already a good equation of
state of isospin symmetric infinite nuclear matter \cite{nucmat,lutzpot}. The
step to inhomogeneous many-nucleon systems is done with the help of the
density-matrix expansion of Negele and Vautherin \cite{negele}. The specific 
isospin-structures of the $1\pi$- and $2\pi$-exchange interaction show up
through relative isospin-factors $-1/3$, $\pm 2/3$ and $-5/3$ of various
subcontributions. We find that the parameterfree results for 
the (density-dependent) strength functions $G_{so}(k_f)$ and $G_J(k_f)$ agree 
fairly well with that of phenomenological Skyrme forces for densities $\rho >
\rho_0/10$. At very low densities a strong variation of the strength functions 
$G_{so}(k_f)$ and $G_J(k_f)$ with density sets in. This has to do with chiral 
singularities $m_\pi^{-1}$ and the presence of two competing small mass scales
$k_f$ and $m_\pi$. The novel density dependencies of $G_{so}(k_f)$ and 
$G_J(k_f)$ as predicted by our parameterfree (leading order) calculation should
be examined in nuclear structure calculations.


\begin{thebibliography}{99}
\bibitem{skyrme} T.H.R. Skyrme, \textit{Nucl. Phys.} \textbf{9} (1959) 615.\vs
\bibitem{sk3} M. Beiner, H. Flocard, N. Van Giai and P. Quentin, \textit{Nucl. 
Phys.} \textbf{A238} (1975) 29.\vs
\bibitem{skm} H. Krivine, J. Treiner and O. Bohigas, \textit{Nucl. Phys.} 
\textbf{A336} (1980) 155.\vs
\bibitem{skmstar} J. Bartel, P. Quentin, M. Brack, C. Guet and H.-B. Hakansson,
\textit{Nucl. Phys.} \textbf{A386} (1982) 79.\vs
\bibitem{sly} E. Chabanat, P. Bonche, P. Haensel, J. Meyer and R. Schaeffer,
\textit{Nucl. Phys.} \textbf{A627} (1997) 710; \textbf{A635} (1998) 231.\vs
\bibitem{reinhard} M. Bender, P.-H. Heenen and P.-G. Reinhard,  {\it Rev. Mod. 
Phys.} {\bf  75} (2003) 121; and references therein.\vs
\bibitem{walecka} B.D. Serot and J.D. Walecka, {\it Int. J. Mod. Phys.} {\bf 
E6} (1997) 515; and references therein.\vs 
\bibitem{ringreview} P. Ring, {\it Prog. Part. Nucl. Phys.} {\bf 37} (1996)
193; P. Ring, Lecture Notes in Physics 581, eds. J.M. Arias and M. Lozana,
Springer Verlag, (2001), page 195; and references therein.\vs
\bibitem{nucmat} N. Kaiser, S. Fritsch and W. Weise, \textit{Nucl. Phys.} 
\textbf{A697} (2002) 255;  \textit{Nucl. Phys.} \textbf{A700} (2002) 343; and 
references therein.\vs
\bibitem{lutzpot} S. Fritsch and N. Kaiser, \textit{Eur. Phys. J.} \textbf{A17}
(2003) in print; nucl-th/0207057.\vs
\bibitem{negele} J.W. Negele and D. Vautherin, {\it Phys. Rev.} {\bf C5} (1972)
1472.\vs
\bibitem{efun} N. Kaiser, S. Fritsch and W. Weise, \textit{Nucl. Phys.} 
\textbf{A} (2003) in print; nucl-th/0212049.\vs
\bibitem{pearson} J.M. Pearson, S. Goriely and M. Samyn, {\it Eur. Phys. J.}
{\bf A15} (2002) 13; F. Tondour, S. Goriely, J.M. Pearson and M. Onsi, {\it
Phys. Rev.} {\bf C62} (2000) 024308.\vs 
\bibitem{isoorbit} P.G. Reinhart and H. Flocard, \textit{Nucl. Phys.} 
\textbf{A584} (1995) 467.\vs
\bibitem{skp} J. Dobaczewski, H. Flocard and J. Treiner, \textit{Nucl. Phys.} 
\textbf{A422} (1984) 103.\vs 
\end{thebibliography}
\end{document}